\documentstyle[prd,eqsecnum,floats,aps]{revtex}
\input psfig.tex
\newcommand{\be}{\begin{equation}}
\newcommand{\ba}{\begin{eqnarray}}
\newcommand{\ee}{\end{equation}}
\newcommand{\ea}{\end{eqnarray}}

\begin{document}
\draft
\def\e{{\bf E}^3}
\def\es{{\bf S}^3}
\def\hy{{\bf H}^3}
\def\hn{{\bf \hat n}}
\def\cp{{\bar C}}
\title{The topology of the universe:  the biggest manifold of them all}

\author{Janna Levin, Evan Scannapieco, and Joseph Silk}

\address{Center for Particle Astrophysics,
UC Berkeley, Berkeley, CA 94720-7304}
\twocolumn

\maketitle
{\widetext

\begin{abstract}

Clues as to the geometry of the 
universe are encoded in the cosmic background radiation.
Hot and cold spots in the primordial radiation may be randomly distributed
in an infinite universe while in a universe with compact topology
distinctive patterns can be generated.
With improved vision, we could actually see if the universe is
wrapped into a hexagonal prism or a hyperbolic horn.  We discuss the 
search for such geometric patterns in predictive maps of the microwave
sky.

\end{abstract}}

\begin{picture}(0,0)
\put(410,170){{ CfPA-98-TH-04}}
\end{picture} \vspace*{-0.15 in}

\vskip 30truept
\centerline{Contribution to the conference proceedings for the Cosmology and 
Topology Workshop}

\vskip 5truept
\centerline{to appear in {\it Class. and Quant. Grav.}}
\twocolumn
\narrowtext


The world abounds with objects that have multiconnected 
topology.  The coffee cup comes to mind as the overused but adored
example of a compact surface with genus 1.  There is also the coffee
cup's topological equivalent, the donut.  People,
buildings, the Earth, are all topologically connected.
As you begin to scan the examples you realize
that every thing is at least finite if not actually
multiply connected.  By contrast, no thing is infinite.
Yet cosmology has persisted in the assumption that the universe
is infinite.  Try explaining that to your friends and neighbors.

Ignoring the global topology of the universe is particularly
neglectful in a theory that purports to be a theory of geometry.  Einstein's
revolution pivots on discarding the notion of a gravitational
force in favor of a theory of geometry.  Yet his theory is incomplete.
While general relativity determines the local curvature of spacetime,
it falls short of specifying the global topology.  A complete theory
of gravity should be able to predict topology \cite{lum}.  
In the absence of a theory of global geometry, we can still wonder how 
natural it would be to live in a given cosmos.
In all,
there are an infinite number of multiconnected topologies
but only three simply connected ones.
Consistent with the belief that 
we are not in a privileged place in the universe, we should not 
live in a geometrically special space.  
The proliferation of compact spaces, particularly hyperbolic, compact
spaces \cite{thur}, indicates that a compact universe may be more probable
\cite{css}.
A quantum cosmological treatment is needed to make this notion more
precise \cite{{gibbons},{carlip}}.
While it is difficult to define a measure on the set of compact spaces,
it is at least clear that
the assumption of an infinite
universe has no basis on which to claim naturalness.

We can observe topology despite our inability to predict it.
Granted, an era of inflation would likely push
the size of the fundamental domain beyond the observable universe.
However, inflation also predicts that the curvature radius is huge
relative to the size of the observable universe.  
Since observational
evidence is favoring a negatively curved space, theoretical prejudices
are considerably weakened.
If curvature is observable, then perhaps so too is topology.

Limits on a universal topology scale are often set by
searching for periodicity in observations of large 
structures \cite{{flat},{rou}}.
Ghost images of any source appear as light wraps around the universe along
different paths and replicates its image for any observer.  
This leads to an extreme version of Obler's paradox whereby a single
bright source could light up the sky if left shining for long enough
\cite{conv}.
On an astronomical scale, there would be ghosts of galaxies and quasars.  
Since
the ghosts are not identical clones, but are rather images of the source
at different ages,
a search for ghosts is impeded by
evolutionary effects and in the end
may be a difficult way to identify the
connectedness of space.

A far more sensitive probe is offered by the
the cosmic background radiation
(CBR).  This
one relatively unmarred fossil record from 
the early universe can reveal the large-scale landscape
of the universe.  
Tiny fluctuations in the spacetime geometry appear as 
hot and cold spots in the otherwise smooth primordial
radiation.
The geometric fluctuations at the time 
light last scatters can be 
reconstructed from these hot and cold spots
and offers a critical
test of small topologies.  
Only perturbations that fit inside the compact
manifold are allowed. The result is a
pattern of fluctuations on the microwave sky reflecting the full structure of
spacetime, local and global. 

The challenge is to predict the spatial pattern given the geometry.  
We discuss the reflection of geometry in the CBR sky in flat 
topologies (\S \ref{compf}) and
in the one multiconnected (but not fully compact) hyperbolic space for which 
CBR maps have been constructed (\S \ref{hornsec}).  
None of the fully compact hyperbolic
geometries have been constrained and have in fact become the topic
of much recent interest
\cite{{lum},{css},{bps},{lbbs}}.
We are learning from  
the simpler spaces how to predict the features in the microwave sky 
for the resistant compact hyperbolic cosmos (\S \ref{comph}).

\section{Compact, Flat Cosmologies}\label{compf}

We want  to predict a
map of the temperature fluctuations.
In a homogeneous and isotropic
space, an angular average over the fluctuations contains all of the
essential information.
Of the six compact, orientable flat spaces, all destroy global isotropy
and all except for the hypertorus destroy global homogeneity.
As a result, there is more information in a map of temperature fluctuations
than just the angular power spectrum.  
Although we argue that the angular average
overlooks conspicuous features in general, for the equal sided
flat cases 
angular spectra do provide a reasonable bound.

Four of the six orientable, compact topologies of $\e$ are constructed from 
a parallelepiped
as the  fundamental domain.  The other two
are built from a hexagonal prism.
The hypertorus is the simplest and has been studied by
many authors \cite{{flat},{moreflat},{sss},{add}}.
Stevens, Scott, and Silk \cite{sss} pointed out that in a flat $3$-torus,
the spectrum of temperature fluctuations 
was truncated at long wavelengths in order to fit within the finite box.
Contrary to standard lore,
we find all of the equal sided compact flat manifolds show a
truncation in the power of fluctuations 
on wavelengths comparable to the size of
the fundamental domain 
\cite{{lss},{tarun}}.
The longest wavelength
fluctuation observed, namely the quadrupole, is in fact low.
Some might even take this as evidence for topology \cite{workshop}.
Cosmic variance is also large on large scales.
Consequently, a fundamental domain the size of the observable universe
is actually consistent with the COBE data \cite{lss}.

A very small universe however 
is
incompatible with the data.
The cutoff in long wavelength perturbations is accompanied by
gaps in power at wavelengths that do not correspond to 
integer windings through the fundamental domain.
{\it All} compact spaces show discrete harmonics and 
as such
the sharp harmonics 
may be a more generic sign of compact topology.
The jaggy
spectra of such small 
compact flat spaces are tens of times less likely than the smooth
spectrum of infinite $\e$.  We conclude, quite conservatively, that
the universe, if finite and flat and equal-sided, must be at least
$80$\% the radius of the
surface of last scatter and so $40$\% of the diameter of the 
observable universe.
There could still be as many as eight
copies of our universe within the observable horizon.

\begin{figure}
\centerline{{\psfig{file=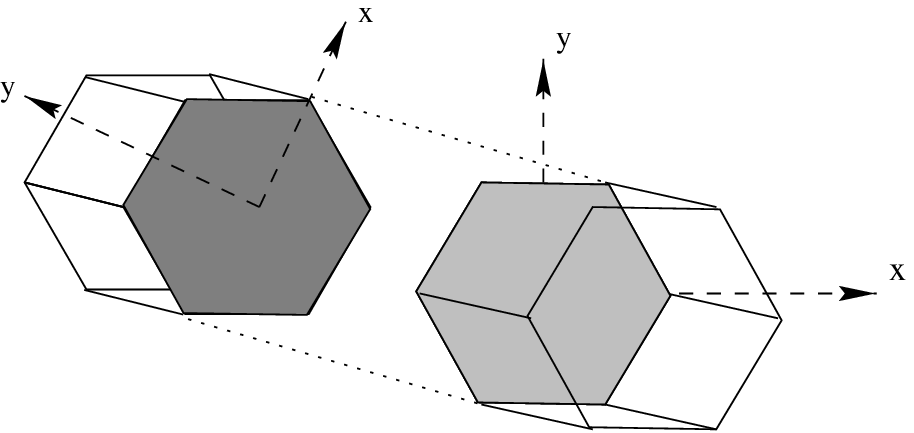,width=3in}}}
\centerline{{\psfig{file=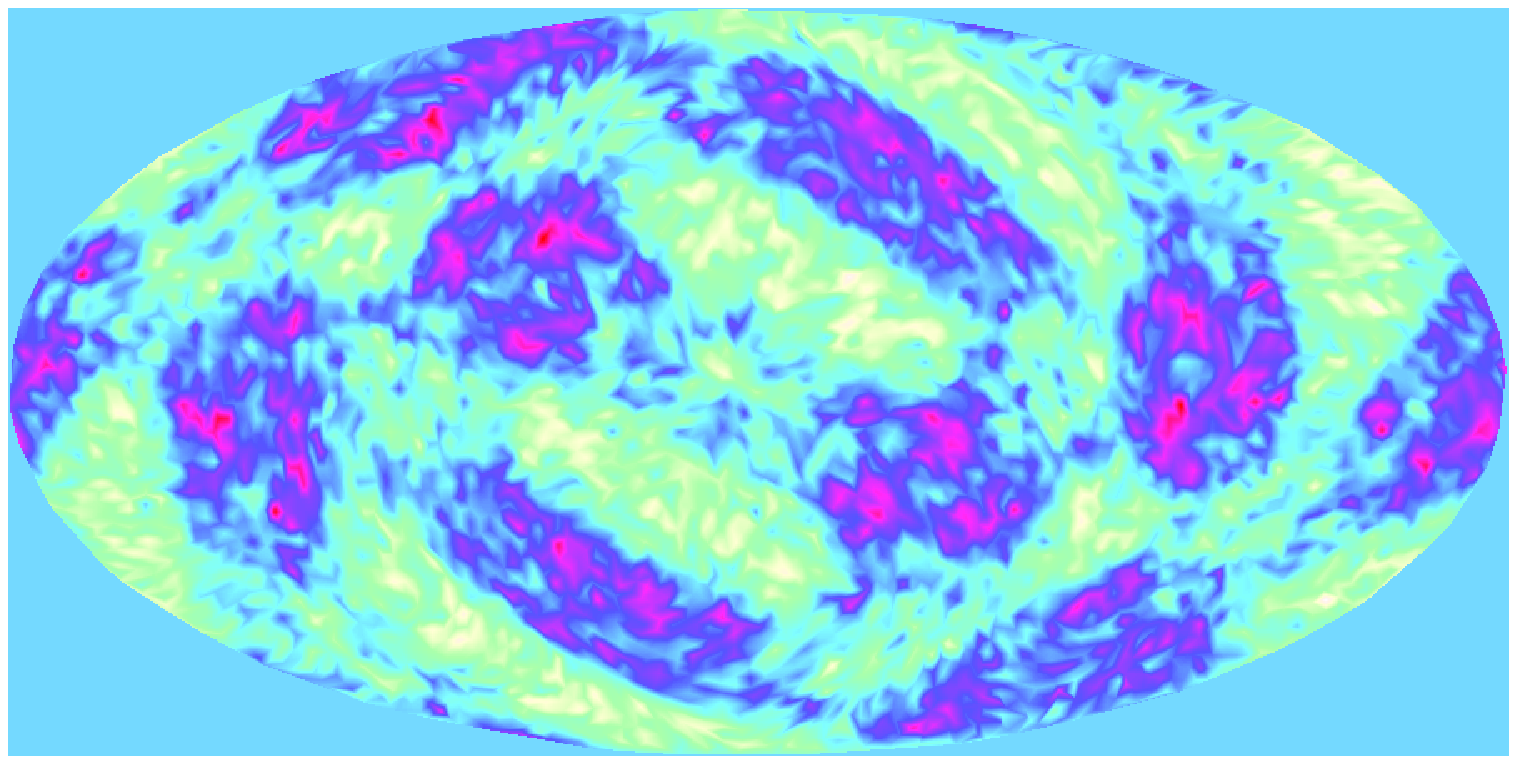,width=3in}}}
\caption{A hexagonal prism with a $2\pi/3$ twist.  The observer is at the
center of the universe in the map of $\delta T(\hat n)/T$.
The fundamental domain is half the diameter of the observable universe
in two directions and one-tenth that in the twisted direction.
\label{fighex}}
\end{figure}

If instead of an equal-sided space we consider a 
fundamental domain
with disparate length scales, the angular power spectrum is 
in general a poor
discriminant. The averaging over the sky fails to recognize the strong
features in the cosmos.
Fig.\ \ref{fighex} shows a predictive map of the hot and cold fluctuations
in a $2\pi/3$-twisted hexagonal prism.
We have set the length of the fundamental domain to be ten times
smaller in the twisted direction than along the face of the hexagon.
The average large angle power in fluctuations is actually consistent 
with the data, although clearly this anisotropic space does not
look like the sky we observe.
A better statistic to discern patterns and correlations is badly needed
\cite{{lss},{lssb}}.
The promising suggestions of \cite{{bps},{css},{lssb}} may be the key
and are discussed more in \S \ref{comph}.

\begin{figure}
\centerline{{\psfig{file=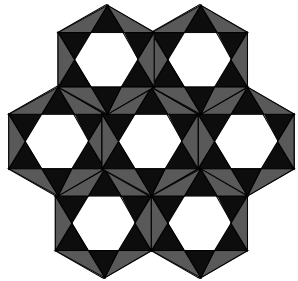,width=1.5in}}}
\centerline{{\psfig{file=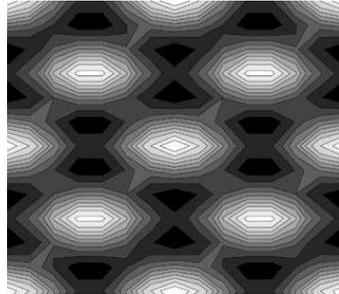,width=1.75in}}}
\caption{Top: A guess at one mode.  Bottom:
A contour plot of the temperature fluctuation for a similar mode.  
\label{onemode}}
\end{figure}

\begin{figure}
\centerline{{\psfig{file=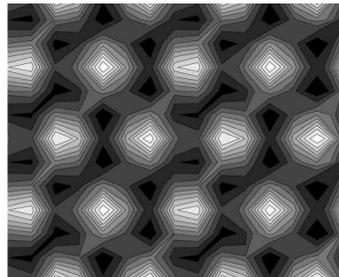,width=1.75in}}}
\caption{A contour plot of the temperature fluctuation for a
mode that winds through the twisted prism.  
\label{mode2}}
\end{figure}

We could have predicted certain features of the map of Fig.\ \ref{fighex},
even if we had not known the eigenmodes explicitly.
A 2D slice through the 3D tiling of space is represented in Fig.\
\ref{onemode}.  If we draw bands connecting opposite sides of the hexagons
and highlight any overlaps, we can predict the imprint of one mode
as shown on the top of Fig. \ref{onemode}.
Given that we do know the eigenmodes, we can show the actual contour plot of 
the hot and cold fluctuations for a similar mode on the bottom of 
Fig.\ \ref{onemode}.  Comparing the guess with the actual contours shows
our guess did quite well.
The hexagonal shape of the universe is clearly seen.
In actuality, there are many modes competing to imprint a pattern on
the sky which blurs the signature hexagons.
In Fig.\ \ref{mode2},
is another contour plot of hot and cold spots for a different mode
which exhibits the $2\pi/3$  twist through the prism.

The competition between fluctuations obscurs some features while enhancing
others.
The surface of last scatter cuts a sphere out of the full 
3D space, an elliptic projection of which is given in the map of 
Fig. \ref{fighex}.
Can you see hexagons in the map?  
Almost.  
Is the $2\pi/3$ twist in the space visible? 
We are currently developing ways of looking at the sky
that pull the underlying patterns out of the noise \cite{lssb}.

\section{Multiconnected, hyperbolic horn }\label{hornsec}

\begin{figure}[h]
\centerline{{\psfig{file=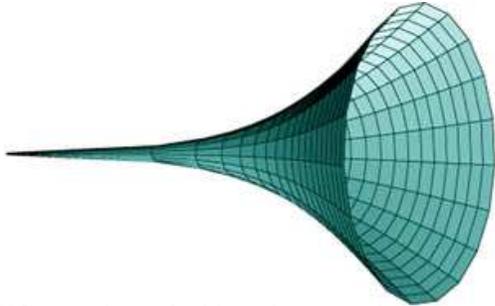,width=3.0in}}}
\caption{
An embedding diagram of the horn with one of the dimensions 
hidden.
\label{embed}}
\end{figure}

No compact hyperbolic manifolds have yet been constrained.
The only multiconnected hyperbolic space for which predictive maps
have been constructed is the horn topology introduced by Sokolov and 
Starobinsky 
\cite{sos}.
The horn is not completely compact but is only closed off in two directions.
A two dimensional subspace is wrapped into a flat torus.  This torus is then
conformally stretched or shrunk along the third dimension tracing out a horn
(Fig.\ \ref{embed}).
There is no chaos on this space since it is only partly compact.
We were thus able to find
exact solutions for the geodesic motions and the perturbation spectrum. 
While the horn topology may seem a special case, there are many manifolds that
may bear horn-like corners.  As emphasized in \cite{css}, many manifolds have
cusps.  To an observer nested in this region the world looks
very much like
a toroidal horn.

\begin{figure}
\centerline{{\psfig{file=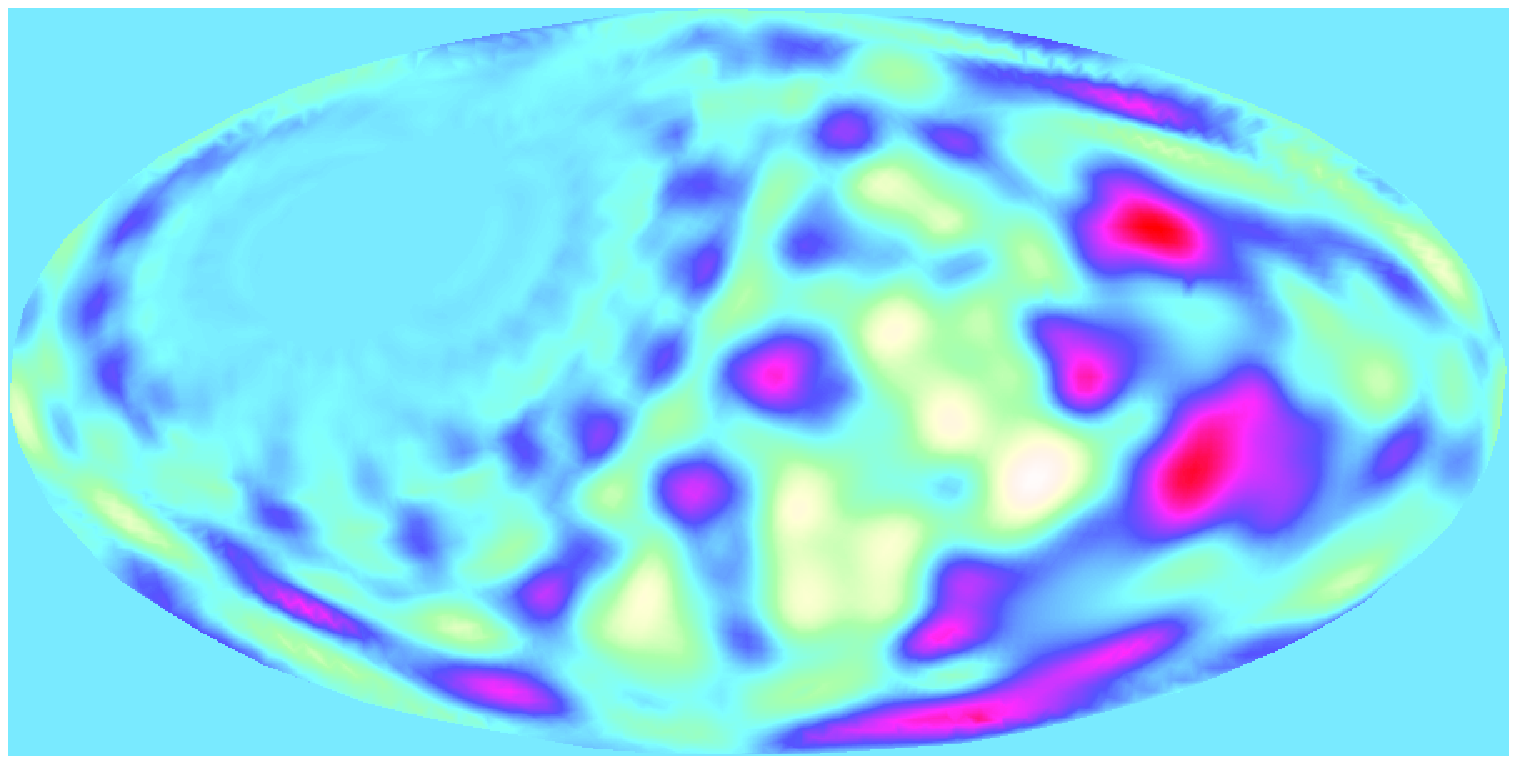,width=3in}}}
\centerline{{\psfig{file=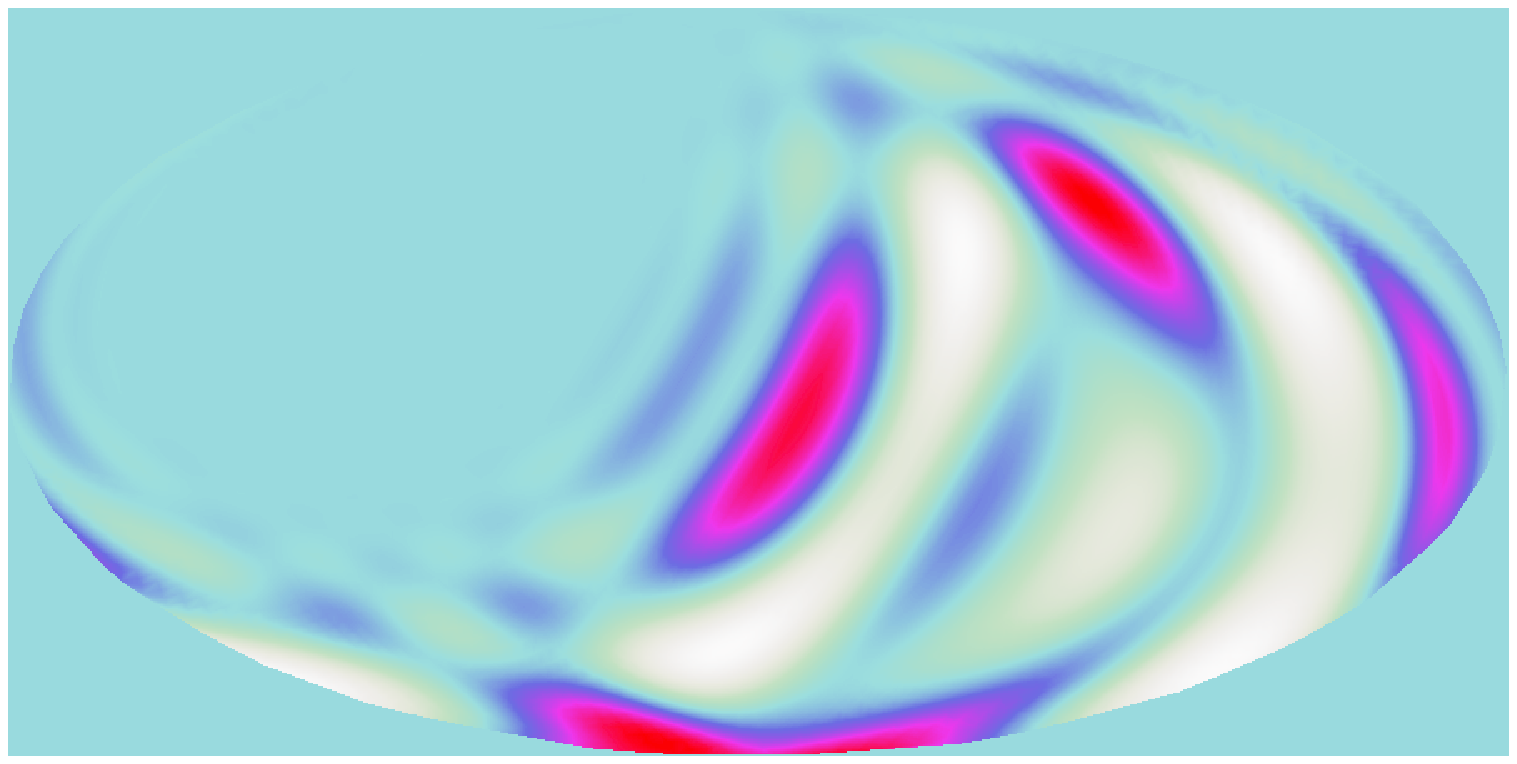,width=3in}}}
\caption{The predicted COBE sky in a horn topology.
\label{horn}}
\end{figure}

In the narrow throat of the horn, big hot and cold spots cannot be 
supported and the temperature of the CBR looks smooth there
as demonstrated in the simulated maps of Fig.\ \ref{horn}.
The upper panel shows a horn with topology scales around
the observer equal to the
radius of the last scattering surface in one direction and half
that in the other compact direction.  
Notice that the observer can see exponentially deep down the 
narrowing throat
of the horn.
The bottom panel shows topology scales equal to $67$\% of the
radius of the last scattering sphere in one compact
direction and
$10$\% in the other.
Again,
you can actually see the geometry of the
horn \cite{lbbs}.

As before, the averaged angular power
spectrum is a weak criterion by which to search for topology.  The 
horn is infinite if multiconnected.  There is therefore formally no 
cutoff in the angular power spectrum.    
Long wavelength modes cannot fit in the 
two compact directions but 
infinitely long modes can exist
along the axis of the horn.
Again, the anisotropy and inhomogeneity would be washed out if we only
considered the angular power spectrum and never looked at a map of the
full $\delta T(\hat n )/T$.
If the universe were completely compact and hyperbolic
with cuspy corners,
there may again be no formal cutoff in long wavelength power \cite{css}.
However, there may still be patches in the sky that were too small
to support fluctuations and hence flat spots would appear.

Looking at the sky as seen by COBE,
it does not appear that 
we live within view of any cuspy corners.
It also appears that
a pattern imprinted on the microwave sky 
will likely be more incisive in identifying topology
than ghost images.  The pattern may not bear a flat
spot, but it will reflect the underlying symmetry group of the
manifold.

\section{Compact, hyperbolic}\label{comph}

In the previous two sections we were able to predict the distribution of 
hot and cold spots across the sky given the geometry. 
We could do so because we could solve for the spectrum of temperature
fluctuations
explicitly.
In a totally compact, hyperbolic space however, the spectrum
cannot be obtained analytically.
The inability to decompose fluctuations in the photon background is a 
direct consequence of the chaotic mixing on compact, negatively curved
surfaces.

Techniques have been attempted in order to circumvent the eigenvalue problem.
The correlation function has been computed using the method
of images \cite{bps}.  
This method involves the sum of the correlation function in the
universal covering space with ghost images in the copies.  The 
temperature 
correlation function can then be compared with the COBE data. 
The method
requires a detailed knowledge of the elements of the symmetry group.  
The sum is highly divergent and is difficult in itself to manage.
The divergence results since 
the number of images proliferates at large distances.

Another approach focuses on the effect of topological lensing of the
last scattering surface.  A well planned statistic may be found which 
can determine if two rings in the sky are really copies of the same
collection of points on the surface of last scatter \cite{{css},{cssthis}}.
High resolution and sensitivity are demanded of an experiment to 
observe the thin rings.
The future satellites such as the Microwave Anisotropy Probe (MAP)
and {\it Planck} are needed
to discriminate a universe with 
such circles in its sky from one that is simply connected
\cite{{ks},{css}}.

Another alternative is to emphasize the spatial patterns in the
sky, although ultimately all the approaches are interwoven.  
In the previous examples,
the universe clearly showed spatial patterns that exposed
the underlying geometry.  
By understanding the symmetries of the fundamental polyhedron
and the identification rules, 
a CBR pattern can be deduced without the need to explicitly 
obtain the spectrum mode by mode \cite{lssb}. 
Given a breakdown of the
pattern, a method not unlike that suggested in \cite{jeff} can be 
implemented to
reconstruct the fundamental domain.  A similar philosophy was used
to 
search for the symmetry axes of a hypertorus in \cite{add}.
No such axis of symmetry was detected, refining 
the bound on anisotropic hypertori.  
 
Geometric patterns have also been used successfully to predict COBE
maps for the Bianchi classes \cite{bian}. 
These Bianchi classes are directly related to Thurston's eight geometries
\cite{thur}.
In addition to the three of constant curvature, there are five more homogeneous
but anisotropic spaces. 
The 
sky patterns can be
predicted just from a knowledge of the group invariances that generate the
geometries and their geodesic flows \cite{{BJS1},{bl}}.
There is a physical difference here from the multiconnected models.
For the Bianchi cosmologies, there are assumed to be no 
initial perturbations.  The
temperature fluctuations in the CBR
result as the shear and expansion of the evolving geometry Doppler shift 
the photons while they transit the
anisotropic space.  

There is even precedence 
in biology.  Spatial pattern formation emerges on the backs of mammals and in
part answers the question of how the leopard got its spots
\cite{murray}.  The enzymes
responsible for the pigmentation on an animals fur fluctuate through
the body.  The geometry of the developing animal and its
scale relative to the characteristic
wavelength of the fluctuating enzymes differentiate the markings.
The universe in its early stages of development similarly acquires
markings.
Again, the geometry and scale of the space
relative to the characteristic
wavelength of the fluctuations in the CBR can influence patterning.
We have already seen the universe wear different coats
in Figs. \ref{fighex} and \ref{horn} reminiscent of a leopard's spots
or a zebra's stripes.  
Surprisingly, biology is a
cleaner system than cosmology
since diffusion mechanisms often single out one mode and thereby
one clean pattern.  In cosmology, many modes are summed and so many
patterns compete in making the universe's coat.
This was evidenced in the hexagons of Fig. \ref{fighex} 
versus those of Figs.\
\ref{onemode} and \ref{mode2}.
It is a theoretical prejudice 
that fundamental physics is simple and everything 
else is just a very messy composite.  Still, the
codes of nature replicate themselves on scales vast and small.
It will be ironic if in this instance we have to first turn to biology to find
the clues to the universe and perhaps even a fundamental theory of gravity.

As the observational evidence accumulates, we are forced
to confront the very real possibility that the universe is negatively
curved and that the curvature scale is just coming within view
with an associated topology scale.
If we do live in a compact, hyperbolic cosmos, we may have to rewrite the
standard big bang story.
A loose narrative for the history of the universe begins with the birth of the
universe and implores an answer from quantum cosmology:  Is the
creation of compact, hyperbolic 3-manifolds favored over others?  Once such a
universe is born, chaotic mixing on the small space could lead to the
relatively smooth cosmos observed today and has already been suggested as an
alternative to inflation
\cite{ellis}.  In fact, it is precisely this suggestion which 
spurred some searches for a small universe.  
While chaotic mixing could explain the average smoothness, the geometry
might explain the local clumpiness of matter.
Pattern formation due to geometry can affect 
the distribution of galaxies as well as the CBR
\cite{{sos},{lbbs}}. 
The global topology
may then create
local inhomogeneity 
and distinctive patterns in the large-scale distribution of luminous and dark
matter. 
The geometry may lead to bubbly or even fractal distributions of 
galaxies.
This scenario for the evolution of our universe is compelling if 
optimistic.  Perhaps geometry alone, and hence the theory of gravity,
determines
the birth and fate of the cosmos.

\section*{Acknowledgment}

We have benefited from valuable discussions with
nearly everyone currently 
active in the area of topology and cosmology.
We are grateful to all of these collaborators
and colleagues: J. Barrow, J. Bond, R. Brooks,
N. Cornish, P. Ferreira, D. Pogosyan, T. Saurodeep, 
D. Spergel, G. Starkman, W. Thurston and 
J. Weeks. 
Special thanks to all of the enthusiastic participants of the 
workshop.



\begin{thebibliography}{99}
\bibitem{tarun}  The CITA group was also aware that the longest modes
were forbidden for some of the twisted flat spaces.  
T. Saurodeep, private communication.
\end{thebibliography}
\end{document}